\documentclass[11pt,onecolumn]{IEEEtran}
\makeatletter
\def\ps@headings{%
\def\@oddhead{\mbox{}\scriptsize\rightmark \hfil \thepage}%
\def\@evenhead{\scriptsize\thepage \hfil \leftmark\mbox{}}%
\def\@oddfoot{}%
\def\@evenfoot{}}
\makeatother
\IEEEoverridecommandlockouts
\usepackage{setspace}
\usepackage[T1]{fontenc}
\usepackage{graphicx,latexsym,subfigure,amsmath,epsfig,latexsym,subfigure,amsmath,cite,amssymb}
\usepackage{algorithm2e,amsthm,epstopdf}
\usepackage{color}
\usepackage{bm}
\usepackage{bm}
\usepackage{url}
\usepackage{multirow}
\usepackage{CJK}
\usepackage{indentfirst}
\usepackage{amsmath}
\usepackage{textcomp}
\usepackage{caption}
\usepackage{cases}
\usepackage{fancyhdr}
\usepackage{mathrsfs}
\usepackage{bm}
\usepackage{float}
\usepackage{setspace}
\usepackage{color}
\usepackage{subfigure}
\usepackage{balance}
\usepackage{subeqnarray}
\usepackage{array}
\usepackage{booktabs}
\usepackage{slashbox}
\usepackage{geometry}
\usepackage{rotating}
 \usepackage{textcomp}
\allowdisplaybreaks[0]

\begin{document}
\title{ IoT Security Techniques Based on Machine Learning}
\author{\footnotesize\IEEEauthorblockN{Liang Xiao\IEEEauthorrefmark{1}\IEEEauthorrefmark{2}, Xiaoyue Wan\IEEEauthorrefmark{1}, Xiaozhen Lu\IEEEauthorrefmark{1},Yanyong Zhang\IEEEauthorrefmark{3}, Di Wu\IEEEauthorrefmark{4}}\\
\IEEEauthorblockA{\footnotesize\IEEEauthorrefmark{1}Dept. of Communication Engineering, Xiamen University, Xiamen, China. Email: lxiao@xmu.edu.cn}\\
\IEEEauthorblockA{\footnotesize\IEEEauthorrefmark{2}National Mobile Communications Research Laboratory, Southeast University, Nanjing, China}\\
\IEEEauthorblockA{\footnotesize\IEEEauthorrefmark{3}WINLAB, Rutgers University, North Brunswick, NJ, USA. Email: yyzhang@winlab.rutgers.edu}\\
\IEEEauthorblockA{\footnotesize\IEEEauthorrefmark{4}Dept. of Computer Science, Sun Yat-Sen University, Guangzhou, China. Email: wudi27@mail.sysu.edu.cn}\\
}

\maketitle
\begin{spacing}{2.0}

\begin{abstract}
Internet of things (IoT) that integrate a variety of devices into networks to provide advanced and intelligent services have to protect user privacy and address attacks such as spoofing attacks, denial of service attacks, jamming and eavesdropping. In this article, we investigate the attack model for IoT systems, and review the IoT security solutions based on machine learning techniques including supervised learning, unsupervised learning and reinforcement learning. We focus on the machine learning based IoT authentication, access control, secure offloading and malware detection schemes to protect data privacy. In this article, we discuss the challenges that need to be addressed to implement these machine learning based security schemes in practical IoT systems.
\end{abstract}

\begin{IEEEkeywords}
IoT security, machine learning, attacks.
\end{IEEEkeywords}

\section{Introduction}\label{1}
Internet of Things (IoT) facilitate integration between the physical world and computer communication networks, and applications (apps) such as infrastructure management and environmental monitoring make privacy and security techniques critical for future IoT systems\cite{Li2011Smart,Sheng2014A,Liu2017A}.
Consisting of radio frequency identifications (RFIDs), wireless sensor networks (WSNs), and cloud computing \cite{Andrea2015Internet}, IoT systems have to protect data privacy and address security issues such as spoofing attacks, intrusions, denial of service (DoS) attacks, distributed denial of service (DDoS) attacks, jamming, eavesdropping, and malwares \cite{Roman2013On,Chen2014A}. 
For instance, wearable devices that collect and send the user health data to the connected smartphone have to avoid privacy information leakage.

It's generally prohibitive for IoT devices with restricted computation, memory, radio bandwidth, and battery resource to execute computational-intensive and latency-sensitive security tasks especially under heavy data streams \cite{Zhou2017Security}. However, most existing security solutions generate heavy computation and communication load for IoT devices, and outdoor IoT devices such as cheap sensors with light-weight security protections are usually more vulnerable to attacks than computer systems.
In this article, we investigate the IoT authentication, access control, secure offloading, and malware detections:
\begin{itemize}
\item[$\bullet$] Authentication helps IoT devices distinguish the source nodes and address the identity based attacks such as spoofing and Sybil attacks \cite{Xiao2016PHY}.
\item[$\bullet$] Access control prevents unauthorized users to access the IoT resources \cite{Abu2014Machine}.
\item[$\bullet$] Secure offloading techniques enable IoT devices to use the computation and storage resources of the servers and edge devices for the computational-intensive and latency-sensitive tasks \cite{Xiao2016A}.
\item[$\bullet$] Malware detection protects IoT devices from privacy leakage, power depletion, and network performance degradation against malwares such as viruses, worms, and Trojans \cite{Xiao2017Cloud}.
\end{itemize}

With the development of machine learning (ML) and smart attacks, IoT devices have to choose the defense policy and determine the key parameters in the security protocols for the tradeoff in the heterogenous and dynamic networks. This task is challenging as an IoT device with restricted resources usually has difficulty accurately estimating the current network and attack state in time. For example, the authentication performance of the scheme in \cite{Xiao2016PHY} is sensitive to the test threshold in the hypothesis test, which depends on both the radio propagation model and the spoofing model. Such information is unavailable for most outdoor sensors, leading to a high false alarm rate or miss detection rate in the spoofing detection.

Machine learning techniques including supervised learning, unsupervised learning, and reinforcement learning (RL) have been widely applied to improve network security, such as authentication, access control, anti-jamming offloading and malware detections
\cite{Abu2014Machine,Ozay2015Machine,Joel2013In,Narudin2016Evaluation,Buczak2016A,Kulkarni2009Neural,
Tan2013A,Xiao2013Proximity,
Xiao2016PHY,Xiao2016A,gwon2013competing,Aref2017Jamming,Li2016SINR,Xiao2017Cloud,Han2017Two}.
\begin{itemize}
\item[$\bullet$] \textbf{Supervised learning} techniques such as support vector machine (SVM), naive Bayes, K-nearest neighbor (K-NN), neural network, deep neural network (DNN) and random forest can be used to label the network traffic or app traces of IoT devices to build the classification or regression model \cite{Abu2014Machine}. For example, IoT devices can use SVM to detect network intrusion \cite{Abu2014Machine} and spoofing attacks \cite{Ozay2015Machine}, apply K-NN in the network intrusion \cite{Joel2013In} and malware \cite{Narudin2016Evaluation} detections, and utilize neural network to detect network intrusion \cite{Buczak2016A} and DoS attacks \cite{Kulkarni2009Neural}. Naive Bayes can be applied by IoT devices in the intrusion detection \cite{Abu2014Machine} and random forest classifier can be used to detect malwares \cite{Narudin2016Evaluation}. IoT devices with sufficient computation and memory resources can utilize DNN to detect spoofing attacks \cite{Shi2017Smart}.
\item[$\bullet$] \textbf{Unsupervised learning} does not require labeled data in the supervised learning and investigates the similarity between the unlabeled data to cluster them into different groups \cite{Abu2014Machine}. For example, IoT devices can use multivariate correlation analysis to detect DoS attacks \cite{Tan2013A} and apply IGMM in the PHY-layer authentication with privacy protection \cite{Xiao2013Proximity}.
\item[$\bullet$] \textbf{Reinforcement learning} techniques such as Q-learning, Dyna-Q, post-decision state (PDS) \cite{He2015Improving} and deep Q-network (DQN) \cite{Mnih2015Human} enable an IoT device to choose the security protocols as well as the key parameters against various attacks via trial-and-error \cite{Xiao2016PHY}. For example, Q-learning as a model free RL technique has been used to improve the performance of the authentication \cite{Xiao2016PHY}, anti-jamming offloading \cite{gwon2013competing,Xiao2016A,Aref2017Jamming}, and malware detections \cite{Li2016SINR,Xiao2017Cloud}. IoT devices can apply Dyna-Q in the authentication and malware detections \cite{Xiao2017Cloud}, use PDS to detect malwares \cite{Xiao2017Cloud} and DQN in the anti-jamming transmission \cite{Han2017Two}.
\end{itemize}

In this article, we briefly review the security and privacy challenges of IoT systems, and investigate the tradeoff between the security performance such as the spoofing detection accuracy and the IoT protection overhead such as the computation complexity, communication latency and energy consumption. We focus on the ML-based authentication, access control, secure offloading, and malware detections in IoT, and discuss the challenges to implement the ML-based security approaches in practical IoT systems.

This article is organized as follows. We review the IoT attack model in Section \ref{2}. We discuss the machine learning based IoT authentication, access control, secure offloading techniques, and malware detections in Sections \ref{3}-\ref{6}, respectively. Finally, we conclude this article and discuss the future work.

\section{IoT Attack model}\label{2}
\begin{figure*}[t]
\centering\includegraphics[width=5.7in]{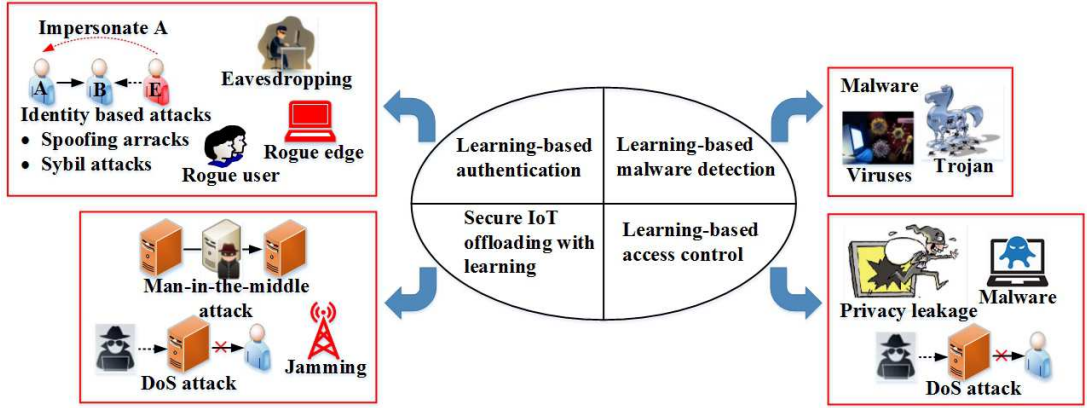}\\
\caption{Illustration of the threat model in Internet of Things.}\label{fig:threat}
\end{figure*}

Consisting of the things, services, and networks, IoT systems are vulnerable to network attacks, physical attacks, software attacks and privacy leakage. In this article, we focus on the IoT security threats as follows.
\begin{itemize}
\item[$\bullet$] \textbf{DoS} attackers aim to prevent IoT devices from receiving the network and computation resources \cite{Andrea2015Internet}.
\item[$\bullet$] \textbf{DDoS} attackers with thousands of IP addresses make it more difficult to distinguish the legitimate IoT device traffic from attack traffic. Distributed IoT devices with light-weight security protocols are especially vulnerable to DDoS attacks \cite{Roman2013On}.
\item[$\bullet$] \textbf{Jamming} attackers send faked signals to interrupt the ongoing radio transmissions of IoT devices and further deplete their bandwidth, energy, central processing units (CPUs) and memory resources of IoT devices or sensors during their failed communication attempts \cite{Han2017Two}.
\item[$\bullet$] \textbf{Spoofing:} A spoofing node impersonates a legal IoT device with its identity such as the medium access control (MAC) address and RFID tag to gain illegal access to the IoT system and can further launch attacks such as DoS and man-in-the-middle attacks \cite{Xiao2016PHY}.
\item[$\bullet$] \textbf{Man-in-the-middle attack:} A Man-in-the-middle attacker sends jamming and spoofing signals with the goal of secretly monitoring, eavesdropping and altering the private communication between IoT devices \cite{Andrea2015Internet}.
\item[$\bullet$] \textbf{Software attacks:} Mobile malwares such as Trojans, worms, and virus can result in the privacy leakage, economic loss, power depletion and network performance degradation of IoT systems \cite{Xiao2017Cloud}.
\item[$\bullet$] \textbf{Privacy leakage:} IoT systems have to protect user privacy during the data caching and exchange. Some caching owners are curious about the data contents stored on their devices and analyze and sell such IoT privacy information. Wearable devices that collect user's personal information such as location and health information have witness an increased risk of personal privacy leakage \cite{Yan2014A}.
\end{itemize}

\begin{table*}[t]
  \small
  \caption{ML-based IoT security methods}
\newcommand{\tabincell}[2]{\begin{tabular}{@{}#1@{}}#2\end{tabular}}
  \centering
  \begin{tabular}{|l|l|l|l|}\hline
Attacks & Security techniques & Machine learning techniques & Performance\\\hline
DoS&\tabincell{l}{Secure IoT offloading\\Access control}&\tabincell{l}{Nerual network \cite{Kulkarni2009Neural}\\Multivariate correlation\\analysis \cite{Tan2013A}\\Q-learning \cite{Li2016SINR}}&\tabincell{l}{Detection accuracy\\Root-mean error}\\\hline
Jamming&\tabincell{l}{Secure IoT offloading}&\tabincell{l}{Q-learning \cite{gwon2013competing,Aref2017Jamming}\\DQN \cite{Han2017Two}}&\tabincell{l}{Energy Consumption\\SINR}\\\hline
Spoofing&\tabincell{l}{Authentication}&\tabincell{l}{Q-learning \cite{Xiao2016PHY}\\Dyna-Q \cite{Xiao2016PHY}\\SVM \cite{Ozay2015Machine}\\DNN \cite{Shi2017Smart}\\Distributed Frank-Wolfe \cite{Yue}\\Incremental aggregated\\gradient \cite{Yue}}&\tabincell{l}{Average error rate\\Detection accuracy\\Classification accuracy\\False alarm rate\\Miss detection rate}\\\hline
Instrusion&\tabincell{l}{Access control}&\tabincell{l}{Support vector machine \cite{Abu2014Machine}\\Naive Bayes \cite{Abu2014Machine}\\ K-NN \cite{Joel2013In}\\ Neural network \cite{Buczak2016A}}&\tabincell{l}{Classification accuracy\\False alarm rate\\Detection rate\\Root mean error}\\\hline
Malware&\tabincell{l}{Malware detection\\Access control}&\tabincell{l}{Q/Dyna-Q/PDS \cite{Xiao2017Cloud}\\Random forest \cite{Narudin2016Evaluation}\\K-nearest neighbors \cite{Narudin2016Evaluation}}&\tabincell{l}{Classification accuracy\\False positive rate\\Ture positive rate\\Detection accuracy\\Detection latency}\\\hline
Eavesdropping&\tabincell{l}{Authentication}&\tabincell{l}{Q-learning \cite{Xiao2016A}\\Nonparametric Bayesian \cite{Xiao2013Proximity}}&\tabincell{l}{Proximity passing rate\\Secrecy data rate}\\\hline
\end{tabular}\label{table}
\end{table*}
\section{Learning-based Authentication}\label{3}
Traditional authentication schemes are not always applicable to IoT devices with limited computation, battery and memory resources to detect identity-based attacks such as spoofing and Sybil attacks. Physical (PHY)-layer authentication techniques that exploit the spatial decorrelation of the PHY-layer features of radio channels and transmitters such as the received signal strength indicators (RSSIs), received signal strength (RSS), the channel impulse responses (CIRs) of the radio channels, the channel state information (CSI), the MAC address can provide light-weight security protection for IoT devices with low computation and communication overhead without leaking user privacy information \cite{Xiao2016PHY}.

PHY-layer authentication methods such as \cite{Xiao2016PHY} build hypothesis tests to compare the PHY-layer feature of the message under test with the record of the claimed transmitter. Their authentication accuracy depends on the test threshold in the hypothesis test. However, it is challenging for an IoT device to choose an appropriate test threshold of the authentication due to radio environment and the unknown spoofing model. As the IoT authentication game can be viewed as a Markov decision process (MDP), IoT devices can apply RL techniques to determine the key authentication parameters such as the test threshold without being aware of the network model.

The Q-learning based authentication as proposed in \cite{Xiao2016PHY} depends on the RSSI of the signals under test and enables an IoT device to achieve the optimal test threshold and improve the utility and the authentication accuracy. For example, the Q-learning based authentication reduces the average authentication error rate by 64.3\% to less than 5\%, and increases the utility by 14.7\% compared with the PHY-authentication with a fixed threshold in an experiment performed in a $12\times9.5\times3$ $\textmd{m}^{3}$ lab with 12 transmitters \cite{Xiao2016PHY}.

\begin{figure*}[t]
  \centering
  \subfigure[Average error rate]{
    \label{fig:subfig:landmark_ave_error}
    \includegraphics[width=2.08in]{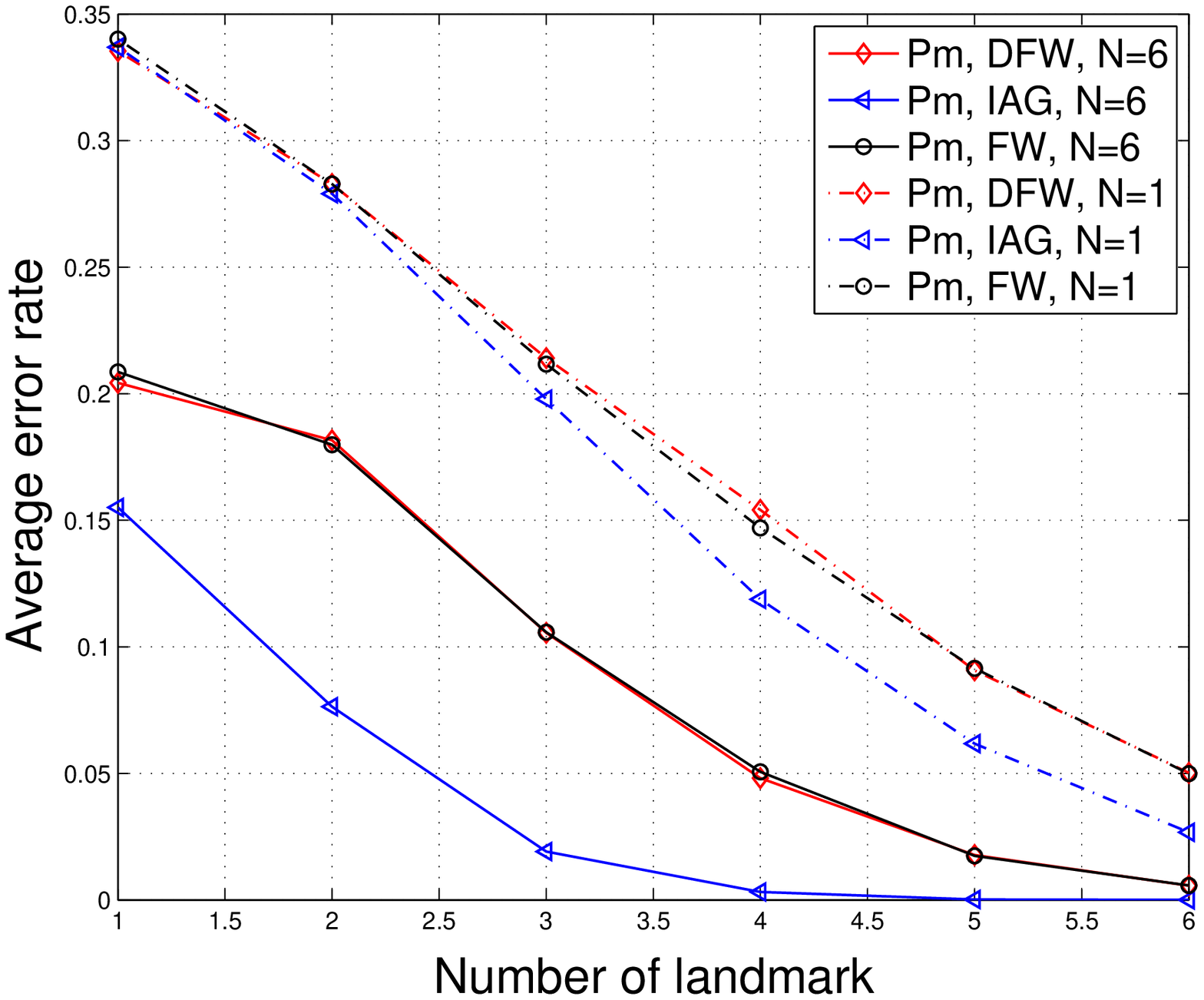}}
    \hspace{-0.3in}
  \subfigure[Communication overhead]{
    \label{fig:subfig:landmark_com_comp_a}
    \includegraphics[width=2.08in]{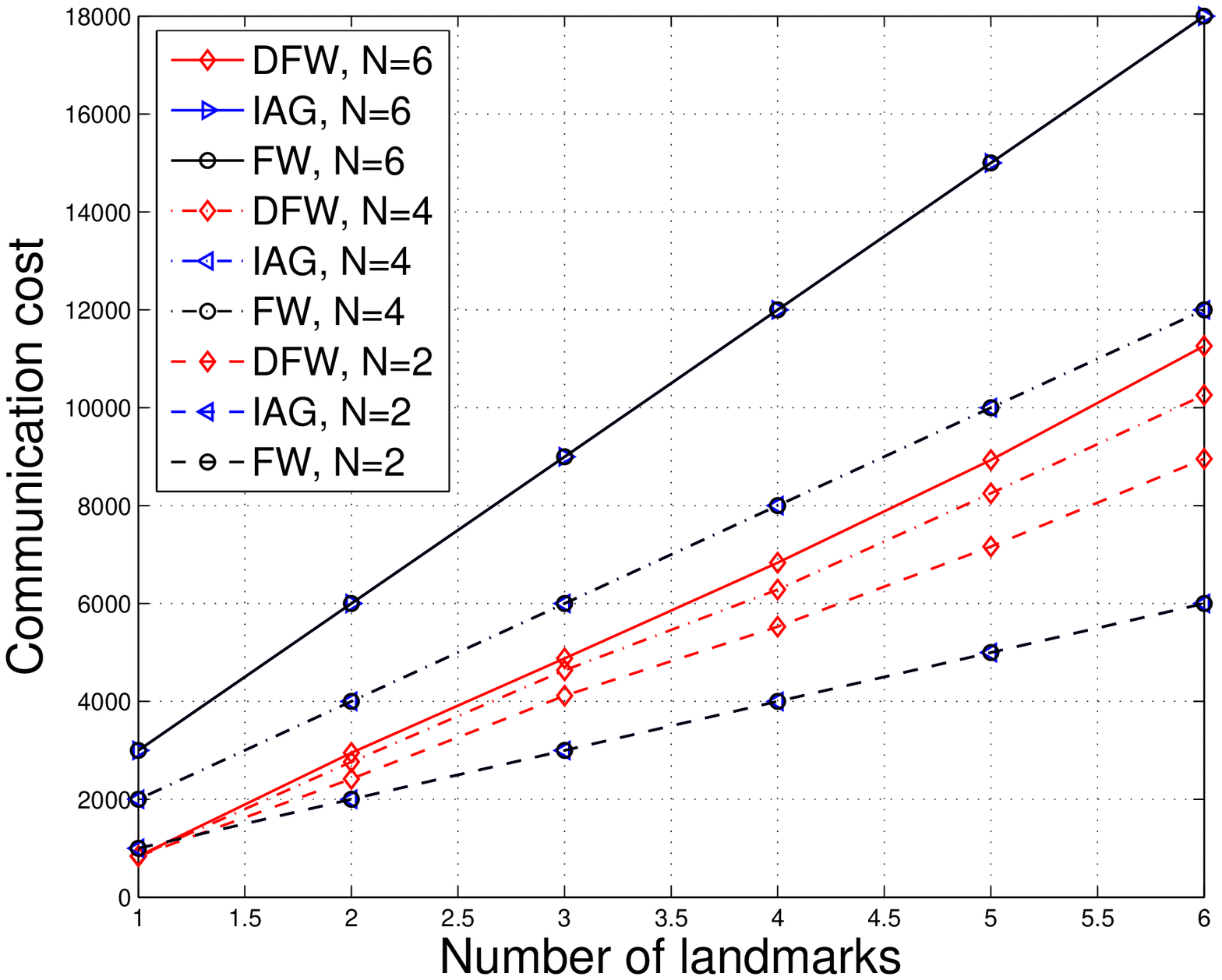}}
  \hspace{-0.3in}
  \subfigure[Computation overhead]{
    \label{fig:subfig:landmark_com_comp_b}
    \includegraphics[width=2.08in]{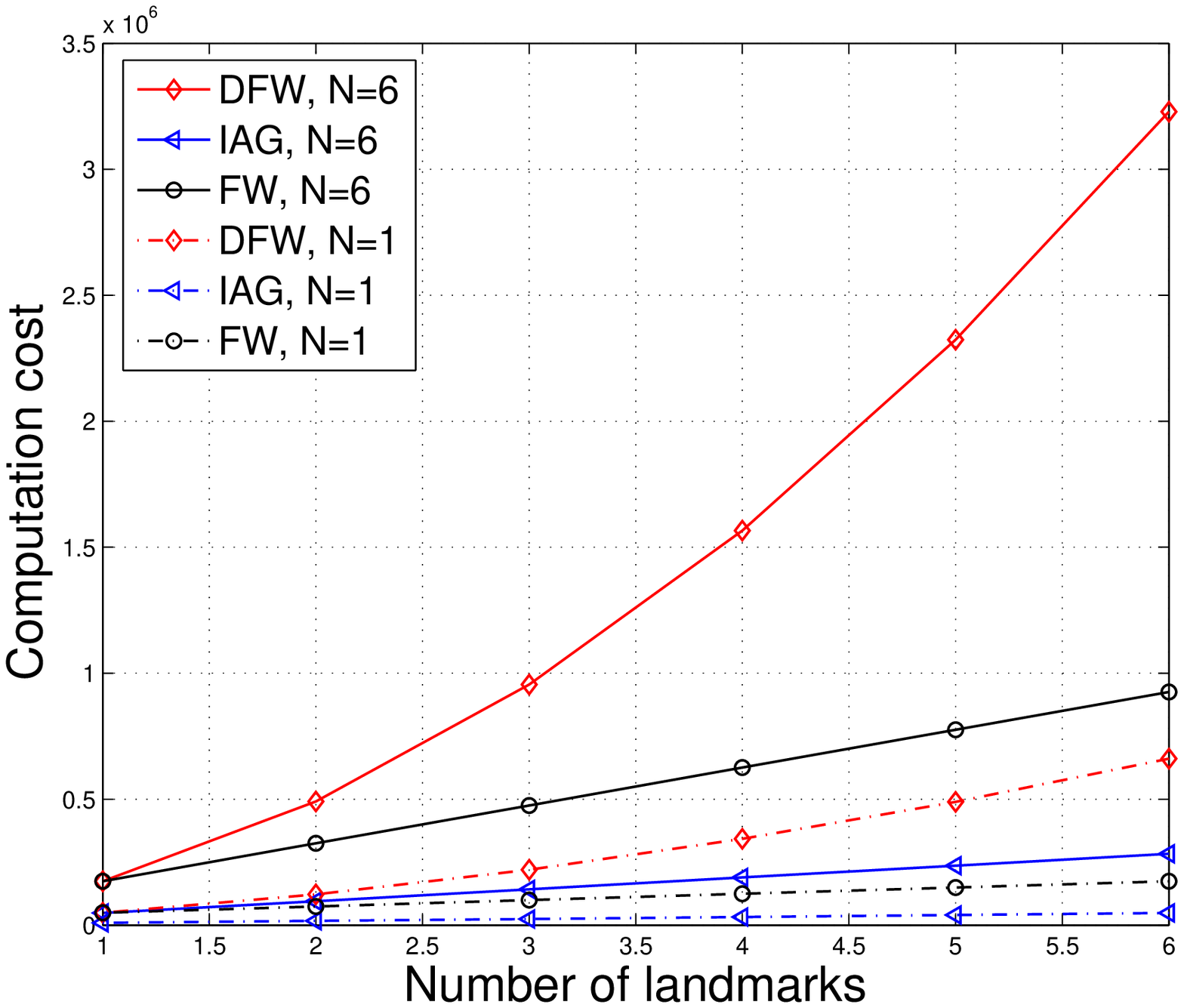}}
  \caption{Performance of PHY-layer authentication system with different number of antennas at each landmark.}
  \label{fig:landmark_performance} 
\end{figure*}

Supervised learning techniques such as Frank-Wolfe (dFW) and incremental aggregated gradient (IAG) can also be applied in IoT systems to improve the spoofing resistance. For example, the authentication scheme in \cite{Yue} applies dFW and IAG and exploits the RSSIs received by multiple landmarks to reduce the overall communication overhead and improve the spoofing detection accuracy. As shown in Fig. \ref{fig:landmark_performance}, the average error rate of the dFW-based authentication and the IAG-based scheme are 6\textperthousand \ and less than $10^{-4}$, respectively, in the simulation with 6 landmarks each equipped with 6 antennas. The dFW-based authentication saves the communication overhead by 37.4\%, while the IAG saves the computation overhead by 71.3 \% compared with the FW-based scheme in this case \cite{Yue}.

Unsupervised learning techniques such as IGMM can be applied in the proximity based authentication to authenticate the IoT devices in the proximity without leaking the localization information of the devices. For instance, the authentication scheme as proposed in \cite{Xiao2013Proximity} uses IGMM, a non-parameteric Bayesian method, to evaluate the RSSIs and the packet arrival time intervals of the ambient radio signals to detect spoofers outside the proximity range. This scheme reduces the detection error rate by 20\% to 5\%, compared with the Euclidean distance based authentication \cite{Xiao2013Proximity} in the spoofing detection experiments in an indoor environment.
\begin{figure*}[t]
\centering\includegraphics[width=5.0in]{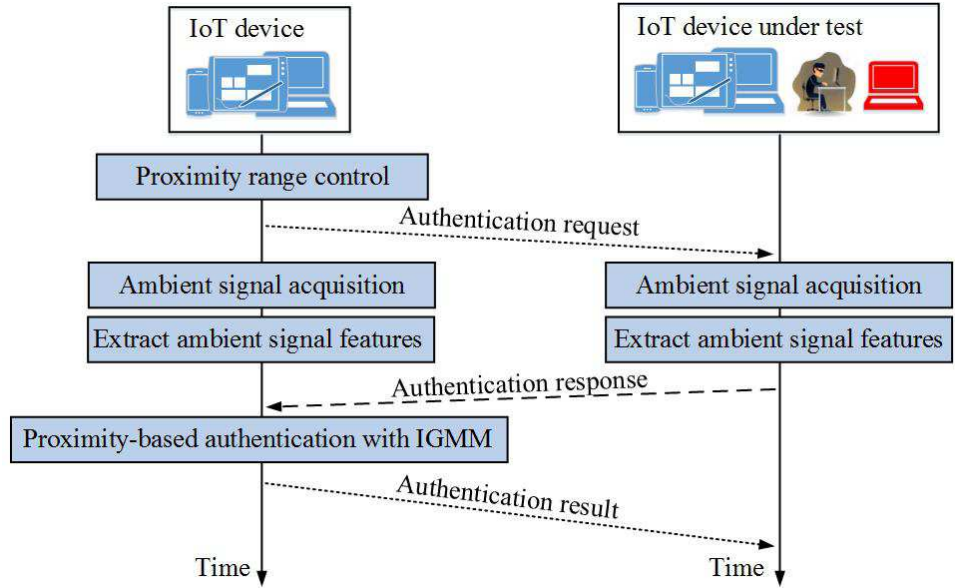}\\
\caption{Illustration of the ML-based authentication in IoT systems.}\label{fig:authentication}
\end{figure*}

As shown in Fig. \ref{fig:authentication}, this scheme requests the IoT device under test to send the ambient signals features such as the RSSIs, MAC addresses and packet arrival time internal of the ambient signals received during a specific time duration. The IoT device extracts and sends the ambient signals features to the legal receiver. Upon receiving such authentication messages, the receiver applies IGMM to compare the reported signal features with those of the ambient signals observed by itself in the proximity based test. The receiver provides the IoT device passing the authentication with access to the IoT resources.

Finally, deep learning techniques such as DNN can be applied for IoT devices with sufficient computation and memory resources to further improve the authentication accuracy.
The DNN-based user authentication as presented in \cite{Shi2017Smart} extracts the CSI features of the WiFi signals and applies DNN to detect spoofing attackers. The spoofing detection accuracy of this scheme is about 95\% and the user identification accuracy is 92.34\% \cite{Shi2017Smart}.

\section{Learning-based Access control}\label{4}
It is challenging to design access control for IoT systems in heterogeneous networks with multiple types of nodes and multi-source data \cite{Abu2014Machine}. ML techniques such as SVM, K-NN  and neural network have been used for intrusion detection \cite{Buczak2016A}. For instance, the DoS attack detection as proposed in \cite{Tan2013A} uses multivariate correlation analysis to extract the geometrical correlations between network traffic features. This scheme increases the detection accuracy by 3.05\% to 95.2\% compared with the triangle area-based nearest neighbors approach with KDD Cup 99 data set \cite{Tan2013A}.

IoT devices such as sensors outdoor usually have strict resource and computation constraints yielding challenges for anomaly intrusion detection techniques usually have degraded detection performance in IoT system. ML techniques help build light-weight access control protocols to save energy and extend the lifetime of IoT systems.
For example, the outlier detection scheme as developed in \cite{Joel2013In} applies K-NN to address the problem of unsupervised outlier detection in WSNs and offers flexibility to define outliers with reduced energy consumption. This scheme can save the maximum energy by 61.4\% compared with the Centralized scheme with similar average energy consumption \cite{Joel2013In}.

The multilayer perceptron (MLP) based access control as presented in \cite{Kulkarni2009Neural} utilizes the neural network with two neurons in the hidden layer to train the connection weights of the MLP and to compute the suspicion factor that indicates whether an IoT device is the victim of DoS attacks. This scheme utilizes backpropagation (BP) that applies the forward computation and error backpropagation and particle swarm optimization (PSO) as an evolutionary computation technique that utilizes particles with adjustable velocities to update the connection weights of the MLP.
The IoT device under test shuts down the MAC layer and PHY-layer functions to save energy and extend the network life, if the output of the MLP exceeds a threshold.

Supervised learning techniques such as SVM are used to detect multiple types of attacks for Internet traffic \cite{Yu2008Traffic} and smart grid \cite{Ozay2015Machine}. For instance, a light-weight attack detection mechanism as proposed in \cite{Yu2008Traffic} uses an SVM-based hierarchical structure to detect the traffic flooding attacks. In the attack experiment, the dataset collector system gathered SNMP MIB data from the victim system using SNMP query messages. Experiment results show that this scheme can achieve attack detection rate over 99.40\% and classification accuracy over 99.53\%, respectively \cite{Yu2008Traffic}.

\section{Secure IoT offloading with Learning}\label{5}
\begin{figure*}[t]
\centering\includegraphics[height=2.1in]{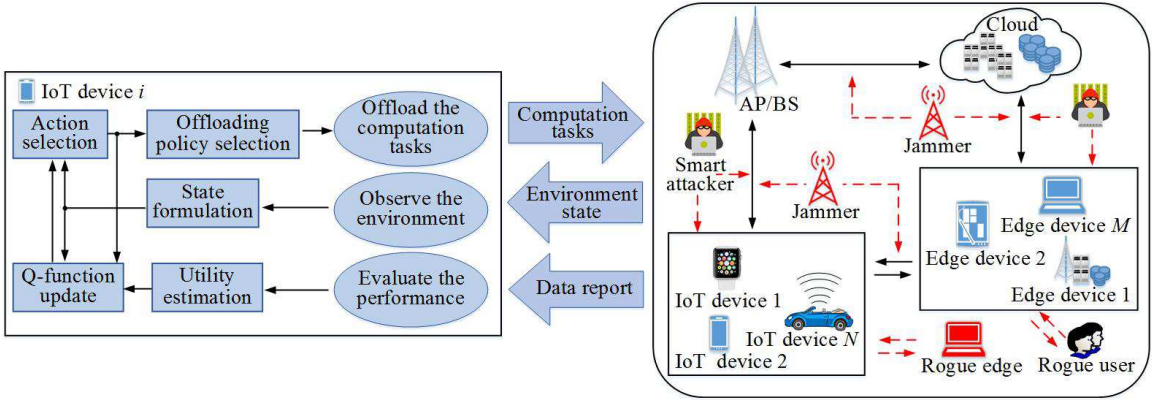}\\
\caption{Illustration of the ML-based offloading.}\label{fig:offloding}
\end{figure*}

IoT offloading has to address the attacks launched from the PHY-layer or MAC layer attacks, such as jamming, rogue edge devices, rouge IoT devices, eavesdropping, man-in-the-middle attacks and smart attacks \cite{Roman2016Mobile}.
As the future state observed by a IoT device is independent of the previous states and actions for a given state and offloading strategy in the current time slot, the mobile offloading strategy chosen by the IoT device in the repeated game with jammers and interference sources can be viewed as a MDP with finite states \cite{Xiao2016A}. RL techniques can be used to optimize the offloading policy in dynamic radio environments.

Q-learning, as a model-free RL technique, is convenient to implement with low computation complexity.
For example, IoT devices can utilize the Q-learning based offloading as proposed in \cite{Xiao2016A} to choose their offloading data rates against jamming and spoofing attacks.
As illustrated in Fig. \ref{fig:offloding}, the IoT device observes the task importance, the received jamming power, the radio channel bandwidth, the channel gain to formulate its current state, which is the basis to choose the offloading policy according to the Q-function. The Q-function, which is the expected discounted long-term reward for each action-state pair and represents the knowledge obtained from the previous anti-jamming offloading.
The Q-values are updated via the iterative Bellman equation in each time slot according to the current offloading policy, the network state and the utility received by the IoT device against jamming.

The IoT device evaluates the signal-to-interference-plus-noise ratio (SINR) of the received signals, the secrecy capacity, the offloading latency the and energy consumption of the offloading process and estimates the utility in this time slot.
The IoT device applies the $\epsilon$-greedy algorithm to choose the offloading policy that maximizes its current Q-function with a high probability and the other policies with a small probability, and thus makes a tradeoff between the exploration and the exploitation. This scheme reduces the spoofing rate by 50\%, and decreases the jamming rate by 8\%, compared with a benchmark strategy as presented in \cite{Xiao2016A}.

According to the Q-learning based anti-jamming transmission as proposed in \cite{gwon2013competing}, an IoT device can apply Q-learning to choose the radio channel to access to the cloud or edge device without being aware of the jamming and interference model in IoT systems. As shown in \ref{fig:offloding}, the IoT device observes the center frequency and radio bandwidth of each channel to formulate the state, and chooses the optimal offloading channel based on the current state and the Q-function. Upon receiving the computation report, the IoT device evaluates the utility and updates the Q values. Simulation results in \cite{gwon2013competing} show that this scheme increases the average cumulative reward by 53.8\% compared with the benchmark random channel selection strategy.

Q-learning also helps IoT devices to achieve the optimal sub-band from the radio spectrum band to resist jamming and interference from other radio devices. As shown in Fig. \ref{fig:offloding}, the IoT device observes the spectrum occupancy to formulate the state and selects the spectrum band accordingly. In an experiment against a sweeping jammer and in the presence of 2 wideband autonomous cognitive radios with 10 sub-bands, this scheme increases the jamming cost by 44.3\% compared with the benchmark sub-band selection strategy in \cite{Aref2017Jamming}.

The DQN-based anti-jamming transmission as developed in \cite{Han2017Two} accelerates the learning speed for IoT devices with sufficient computation and memory resources to choose the radio frequency channel. This scheme applies the convolutional neural network to compress the state space for the large scale networks with a large number of IoT devices and jamming policies in a dynamic IoT systems and thus increase the SINR of the received signals. This scheme increases the SINR of the received signals by 8.3\% and saves 66.7\% of the learning time compared with the Q-learning scheme in the offloading against jamming attacks\cite{Han2017Two}.

\section{Learning-based IoT Malware detection}\label{6}
\begin{figure*}[t]
\centering\includegraphics[height=2.7in]{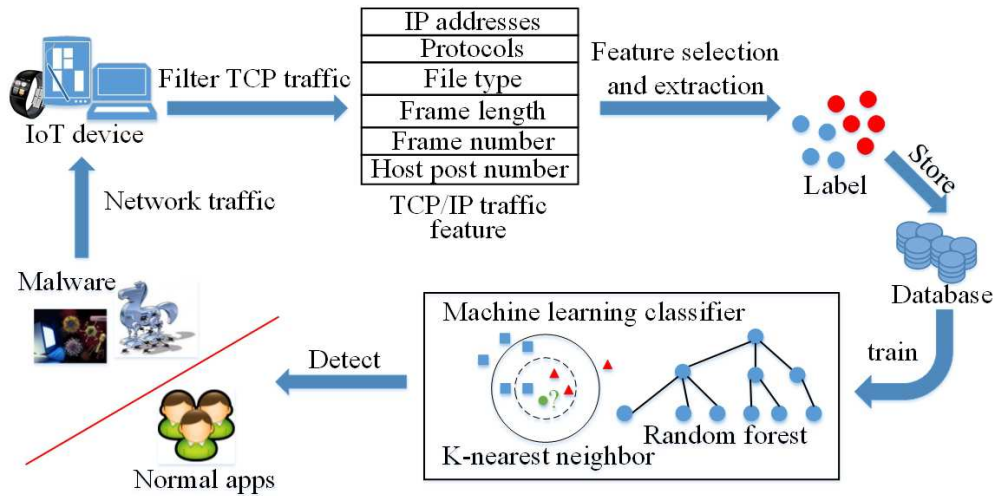}\\
\caption{Illustration of the ML-based malware detection.}\label{fig:malware_detection}
\end{figure*}
IoT devices can apply supervised learning techniques to evaluate the runtime behaviors of the apps in the malware detection. In the malware detection scheme as developed in \cite{Narudin2016Evaluation}, an IoT device uses K-NN and random forest classifiers to build the malware detection model. As illustrated in Fig. \ref{fig:malware_detection}, the IoT device filters the TCP packets and selects the features among various network features including the frame number and length, labels them and stores these features in the database. The K-NN based malware detection assigns the network traffic to the class with the largest number of objects among its $K$ nearest neighbors. The random forest classifier builds the decision trees with the labeled network traffic to distinguish malwares.
According to the experiments in \cite{Narudin2016Evaluation}, the true positive rate of the K-NN based malware detection and random forest based scheme with MalGenome dataset are 99.7\% and 99.9\%, respectively.

IoT devices can offload app traces to the security servers at the cloud or edge devices to detect malwares with larger malware database, faster computation speed, larger memories, and more powerful security services. The optimal proportion of the apps traces to offload depends on the radio channel state to each edge device and the amount of the generated app traces. RL techniques can be applied for an IoT device to achieve the optimal offloading policy in a dynamic malware detection game without being aware of the malware model and the app generation model \cite{Xiao2017Cloud}.

\begin{figure*}[t]
\centering\includegraphics[height=3.2in]{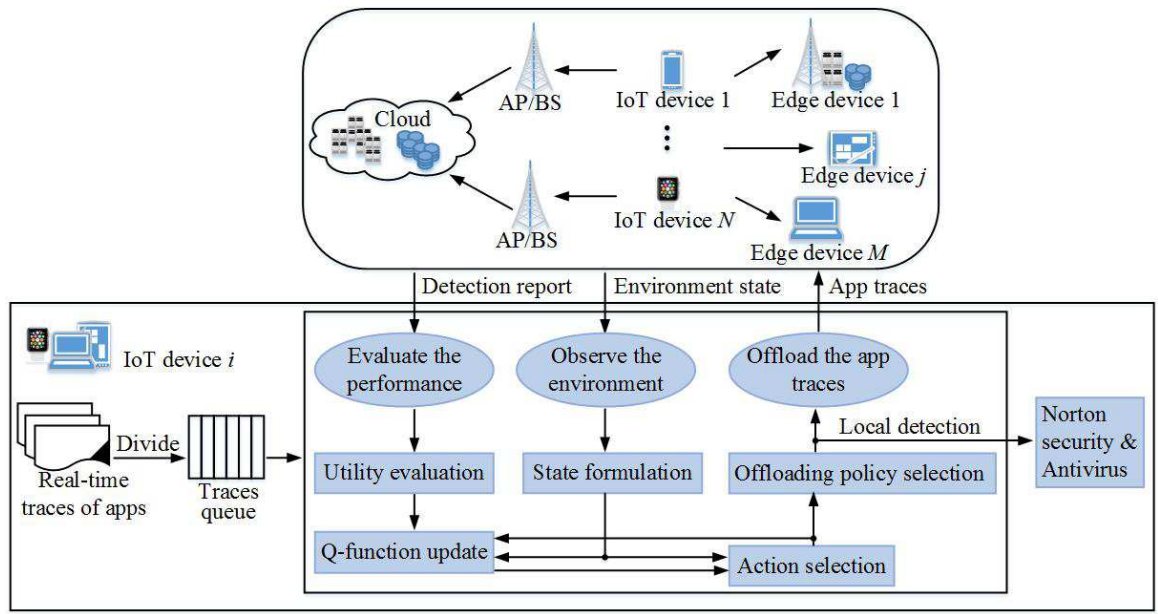}\\
\caption{Illustration of the ML-based malware detection with offloading.}\label{fig:malware_offload}
\end{figure*}

In a malware detection scheme as developed in \cite{Xiao2017Cloud}, an IoT device can apply the Q-learning to achieve the optimal offloading rate without knowing the trace generation and the radio bandwidth model of the neighboring IoT devices. As shown in Fig. \ref{fig:malware_offload}, the IoT device divides real-time app traces into a number of portions, and observes the user density and radio channel bandwidth to formulate the current state. The IoT device estimates the detection accuracy gain, the detection latency and energy consumption to evaluate the utility received in this time slot. This scheme improves the detection accuracy by 40\%, reduces the detection latency by 15\%, and increases the utility of the mobile devices by 47\%, compared with the benchmark offloading strategy in \cite{Xiao2017Cloud} in a network consisting of 100 mobile devices.

The Dyna-Q based malware detection scheme as presented in \cite{Xiao2017Cloud} exploits the Dyna architecture to learn from hypothetical experience and find the optimal offloading strategy. This scheme utilizes both the real defense experiences and the virtual experiences generated by the Dyna architecture to improve the learning performance. For instance, this scheme reduces the detection latency by 30\% and increases the accuracy by 18\%, compared with the detection with Q-learning \cite{Xiao2017Cloud}.

To address the false virtual experiences of Dyna-Q especially at the beginning of the learning process, the PDS-based malware detection schemes as developed in \cite{Xiao2017Cloud} utilizes the known radio channel model to accelerate the learning speed. This scheme applies the known information regarding the network, attack and channel models to improve the exploration efficiency and utilizes Q-learning to study the remaining unknown state space. This scheme increases the detection accuracy by 25\% compared with the Dyna-Q based scheme in a network consisting of 200 mobile devices \cite{Xiao2017Cloud}.

\section{Conclusion and Future Work}\label{Coclusion}
In this article, we have identified the IoT attack models and the learning based IoT security techniques, including the IoT authentication, access control, malware detections and secure offloading, which are shown to be promising to protect IoTs. Several challenges have to be addressed to implement the learning based security techniques in practical IoT systems:
\begin{itemize}
\item[$\bullet$] \textbf{Partial state observation:} Existing RL-based security schemes assume that each learning agent knows the accurate state and evaluate the immediate reward for each action in time. In addition, the agent has to tolerant the bad strategies especially at the beginning of the learning process. However, IoT devices usually have difficulty estimating the network and attack state accurately, and has to avoid the security disaster due to a bad policy at the beginning of the learning process. A potential solution is transfer learning \cite{Pan2010A} that explores existing defense experiences with data mining to reduce the random exploration, accelerates the learning speed and decreases the risks of choosing bad defense policies at the beginning of the learning process. In addition, backup security mechanisms have to be provided to protect IoT systems from the exploration stage in the learning process.
\item[$\bullet$] \textbf{Computation and communication overhead:} However, many existing ML-based security schemes have intensive computation and communication costs, and require a large amount of training data and complicated features extraction process \cite{Abu2014Machine}. Therefore, new ML techniques with low computation and communication overhead such as dFW have to be investigated to enhance security for IoT systems, especially for the scenarios without the cloud-based servers and edge computing.
\item[$\bullet$] \textbf{Backup security solutions:} The RL-based security methods have to explore the "bad" security policy that sometimes can cause network disaster for IoT systems at the beginning stage of learning to achieve the optimal strategy. The intrusion detection schemes based on unsupervised learning techniques sometimes have miss detection rates that are not negligible for IoT systems. Supervised and unsupervised learning sometimes fails to detect the attacks due to oversampling, insufficient training data and bad feature extraction. Therefore, backup security solutions have to be designed and incorporated with the ML-based security schemes to provide reliable and secure IoT services.
\end{itemize}

\section*{Acknowledgment}
This work was supported by the National Natural Science Foundation of China under Grant 61572538 and 91638204, the Fundamental Research Funds for the Central Universities under Grant 17LGJC23, and the open research
fund of National Mobile Communications Research Laboratory, Southeast University (No.2018D08).

\newpage
\bibliography{reference}
\bibliographystyle{IEEEtr}

\begin{IEEEbiographynophoto}
{Liang Xiao}
(M'09, SM'13) is currently a Professor in the Department of Communication Engineering, Xiamen University, Fujian, China. She has served as an associate editor of IEEE Trans. Information Forensics and Security and guest editor of IEEE Journal of Selected Topics in Signal Processing. She is the recipient of the best paper award for 2016 INFOCOM Big Security WS and 2017 ICC. She received the B.S. degree in communication engineering from Nanjing University of Posts and Telecommunications, China, in 2000, the M.S. degree in electrical engineering from Tsinghua University, China, in 2003, and the Ph.D. degree in electrical engineering from Rutgers University, NJ, in 2009. She was a visiting professor with Princeton University, Virginia Tech, and University of Maryland, College Park.
\end{IEEEbiographynophoto}

\begin{IEEEbiographynophoto}
{Xiaoyue Wan}
(S'16) received the B.S. degree in communication engineering from Xiamen University, Xiamen, China, in 2016. She is currently pursuing the M.S. degree with the Department of Communication Engineering, Xiamen University, Xiamen, China.
\end{IEEEbiographynophoto}

\begin{IEEEbiographynophoto}
{Xiaozhen Lu}
(S'17) received the B.S. degree in communication engineering from Nanjing University of Posts and Telecommunications, Nanjing, China, in 2017. She is currently pursuing the PhD. degree with the Department of Communication Engineering, Xiamen University, Xiamen, China.
\end{IEEEbiographynophoto}
\begin{IEEEbiographynophoto}
{Yanyong Zhang}
(M'08, SM'15) received the BS degree in computer science from the University of Science and Technology of China in 1997 and the PhD degree in computer science and engineering from Penn State University in 2002. She joined the Electrical and Computer Engineering Department of Rutgers University as an assistant professor in 2002. In 2008, she was promoted to associate professor with tenure, and in 2015, was promoted to full professor. She is also a member of the Wireless Information Networking Laboratory (Winlab). During March-July 2009, she was a visiting scientist at Nokia Research Center Beijing. Dr. Zhang is the recipient of the US NSF CAREER award. She is currently an associate editor for the IEEE Transactions on Mobile Computing and IEEE Transactions on Services Computing. She has served on TPC for many conferences, including INFOCOM, ICDCS, DSN, IPSN, etc. She is a fellow of IEEE.
\end{IEEEbiographynophoto}
\begin{IEEEbiographynophoto}{Di Wu}
(M'06-SM'17) received the B.S. degree from the University of Science and Technology of China, Hefei, China, in 2000, the M.S. degree from the Institute of Computing Technology, Chinese Academy of Sciences, Beijing, China, in 2003, and the Ph.D. degree in computer science and engineering from the Chinese University of Hong Kong, Hong Kong, in 2007. He was a Post-Doctoral Researcher with the Department of Computer Science and Engineering, Polytechnic Institute of New York University, Brooklyn, NY, USA, from 2007 to 2009, advised by Prof. K. W. Ross. Dr. Wu is currently a Professor and the Assistant Dean of the School of Data and Computer Science with Sun Yat-sen University, Guangzhou, China.
\end{IEEEbiographynophoto}
\balance
\end{spacing}
\end{document}